# Using Software Categories for the Development of Generative Software

Pedram Mir Seyed Nazari[1], Bernhard Rumpe[1]

[1]*Software Engineering, RWTH Aachen University, Germany*
*{nazari, rumpe}@se-rwth.de*

Abstract: In model-driven development (MDD) software emerges by systematically transforming abstract models to concrete source code. Ideally, performing those transformations is to a large extent the task of code generators. One approach for developing a new code generator is to write a reference implementation and separate it into handwritten and generatable code. Typically, the generator developer manually performs this separation — a process that is often time-consuming, labor-intensive, difficult to maintain and may produce more code than necessary. Software categories provide a way for separating code into designated parts with defined dependencies, for example, "Business Logic" code that may not directly use "Technical" code. This paper presents an approach that uses the concept of software categories to semi-automatically determine candidates for generated code. The main idea is to iteratively derive the categories for uncategorized code from the dependencies of categorized code. The candidates for generated or handwritten code finally are code parts belonging to specific (previously defined) categories. This approach helps the generator developer in finding candidates for generated code more easily and systematically than searching by hand and is a step towards tool-supported development of generative software.

## 1 INTRODUCTION

Models are at the center of the model-driven development (MDD) approach. They abstract from technical details, facilitating a more problem-oriented development of software. In contrast to conventional general-purpose languages (GPL, such as Java or C), the language of models is limited to concepts of a specific domain, namely, a domain-specific language (DSL). To obtain an exectuable software application, code generators systematically transform the abstract models to instances of a GPL (e.g., classes of Java). However, code generators are software themselves and need to be developed as well. There are different development processes for code generators. One that is often suggested (e.g., (Kelly and Tolvanen, 2008) and (Schindler, 2012)) is shown in Fig. 1.

The approach includes four steps. First, a reference model is created, which ultimately serves as input for the generator. Depending on this reference model, the generator developer creates the reference implementation. Next, it has to be determined which code parts need to be or can be generated and which ones should remain handwritten. Finally, the transformations are defined to transform the reference model to the aforementioned generated code.

Often, the third step, i.e., 'separation of handwritten and generated code' is not explicitly mentioned in the literature. This separation is implicit part of the last step, i.e., 'creation of transformations', since the transformations are only created for code that ought to be generated. However, the separation of handwritten and generated code ought to be distinguished as a step on its own, since it is not always obvious which classes need to be generated.

In general, every class can be generated, especially when using template-based generators. In an extreme case, a class can be fully copied into a template containing only static template code (and, thus, is independent of the input model). This is not desired, following the guideline that only as much code should be generated as necessary (Stahl et al., 2006), (Kelly and Tolvanen, 2008), (Fowler, 2010). Optimally, most code is put into the domain framework (or domain platform), increasing the understandabil-ity and maintainability of the software. The generated code then only configures the domain framework for specific purposes (Rumpe, 2012).

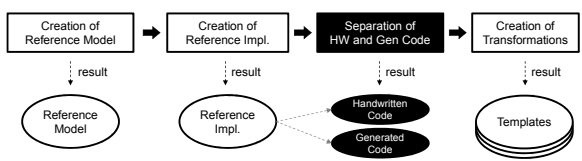

Figure 1: Typical development steps of a code generator.

One important criterion for a code generator to be reasonable is the existence of similar code parts, either in the same software product or in different products (e.g., software product lines). Typically, generation candidates are similar code parts that are also related to the domain. For example, in a domain about cars, the classes `Wheel` and `Brake` would be more likely generation candidates than the domain independent and thus unchanged class `File`. This, of course, is the case, since the information for the generated code is obtained by the input model which, in turn, is an instance of a DSL that by definition describes elements of a specific domain. Of course, the logical relation to the domain is not a necessary criterion, because if the DSL is not expressive enough, the generated code is additionally integrated with handwritten code. Nevertheless, the generated code often has some bearing on the domain.

In most cases, the generator developer manually separates handwritten code from generated code. This process can be time-consuming, labor-intensive and may impede maintenance. Furthermore, when using a domain framework, this separation is insufficient, since the handwritten code needs to be separated into handwritten code for a *specific project* and handwritten code concerning the *whole domain*. This separation also impacts the maintenance of the software (Stahl et al., 2006). To address this problem, software categories, as presented in (Siedersleben, 2004), are suited.

The aim of this paper is to show how software categories can be exploited to categorize semi-automatically classes and interfaces of an object-oriented software system. The resulting categorization can be used for determining candidates for generated code, supporting the developer performing this separation task.

This paper is structured as follows: Sec. 2 introduces software categories and the used terminology. In Sec. 3, these software categories are adjusted for generative software. Sec. 4 presents the allowed dependencies derived by the previously defined software categories. The general categorization approach is explained in Sec. 5 and exemplified in Sec. 6. Sec. 7 outlines further possible dependencies. Finally, Sec. 8 concludes the paper.

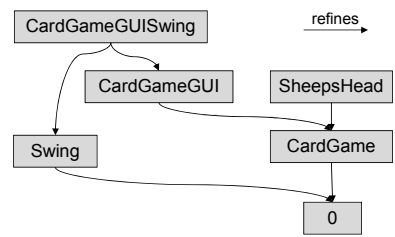

Figure 2: Software categories for virtual *SheepsHead* (Siedersleben, 2004) (shortened).

## 2 SOFTWARE CATEGORIES

Software systems, especially larger ones, consist of a number of components that interact with each other. The components usually belong to different kinds of categories, such as *persistence*, *gui* and *application*. Therefore, (Siedersleben, 2004) suggests using software categories for finding appropriate components. In the following this idea is demonstrated by an example.[1]

Suppose that a software system for the card game Sheepshead should be developed. The following categories then could be created (see Fig. 2):

- *0* (Zero): contains only global software that is well-tested, e.g., `java.lang` and `java.util` of the JDK.

- *CardGame*: contains fundamental knowledge about card games in general. Hence, it can be used for different card games.

- *SheepsHead*: Contains rules for the the Sheepshead game, e.g., whether a card can be drawn.

- *CardGameGUI*: determines the design of the card game, independent of the used library, e.g., that the cards should be in the middle of the screen.

- *CardGameGUISwing*: extends Swing by illustration facilities for cards.

- *Swing*: contains fundamental knowledge about Java Swing.

An arrow in Fig. 2 represents a refinement relation between two categories. Classes that are in a category *C1* that refines another category *C2* may use classes of this category *C2*. The other way around is not allowed. Every category - directly or indirectly - refines the category *0* (arrows in Fig. 2). Hence, software in *0* can be used in every category without any problems. *CardGame* is refined by *SheepsHead* and *CardGameGUI* which means that code in these categories can also use code in *CardGame*. Note that a communication between *CardGameGUI* and *SheepsHead* is not allowed directly, but rather by using *CardGame* or *0* interfaces. Since the category *CardGameGUISwing* refines both *CardGameGUI* and *Swing*, it is a mixed form of these two categories.

---
[1]The example is taken from (Siedersleben, 2004) and reduced to only the aspects required to explain our approach.

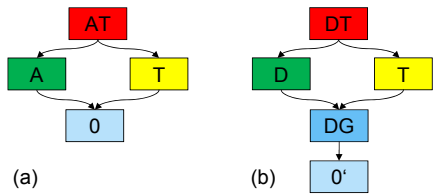

Figure 3: Software categories (a) in general (Siedersleben, 2004) and (b) adjusted for generative software.

Now, having these categories, appropriate components can be found. For example, a component `SheepsHeadRules` in the category *SheepsHead*, `CardGameInfo` and `VirtualPlayer` in *CardGame*, `CardGameInfoPresentation` for *CardGameGUI*.

Considering this example, it can be seen that beside the *0* category, three other categories can be identified that exist in most software systems (see Fig. 3a):

- *Application* (*A*): containing only application software, i.e., *CardGame*, *SheepsHead* and *CardGameGUI*

- *Technical* (*T*): containing only technical software, e.g. Java *Swing* classes.[2]

- Combination of *A* and *T* (*AT*): e.g. *CardGameGUI-Swing* because it refines both an *A* (*CardGameGUI*) and a *T* (`Swing`) category.[3]

(Siedersleben, 2004) summarizes the characteristics and rules for the software categories as follows: the categories are partially ordered, i.e., every category can refine one or more categories. The emerging category graph is acyclic. The category *0* (Zero) is the root category, containing global software. A category *C* is pure, if there is only one path from *C* to *0*. Otherwise, the category is impure. In Fig. 3a only the category *AT* is impure, because it refines the two categories *A* and *T*. All other categories are pure.

---

[2]Note that `Swing` classes are global (belonging to the JDK) and well-tested; hence meet the criteria of the category *0*. But –as usually the user-interface should be exchangeable– `Swing` classes are not necessarily global in a specific software system.

[3]In (Siedersleben, 2004) also the *Representation* (*R*) category is presented. This category contains only software for transforming *A* category software to *T* and vice versa. It is a kind of cleaner version of *AT*. To demonstrate our approach, the *R* category can be neglected.

## Terminology

We call a class that has the category *C* a *C*-class. Following from the category graph in Fig. 2 there are: *AT*-classes, *A*-classes, *T*-classes and *0*-classes. For the sake of readability, we do not explicitly mention interfaces, albeit what applies to classes applies to interfaces as well.

## 3 CATEGORIES FOR GENERATIVE SOFTWARE

While (Siedersleben, 2004) aims for finding components from the defined software categories, the goal of this paper is to determine whether a specific class should be *generated or not* by analyzing its dependencies to other classes.

To illustrate this, consider the following example. When having a class `Book` and a class `Jupiter`, which of these classes are generation candidates? Of course, it *depends on the domain*. If the domain is about planets, probably `Jupiter` is a candidate. In a carrier media domain, `Book` would be a candidate. So, we can say, that a generation candidate somehow relates to the domain. But this condition is not enough. In a library domain where different books exist, `Book` would rather be general for the *whole domain* and should probably not be generated at all. Hence, additionally to the domain affiliation, a generation candidate *is not general for the whole domain*. Technically speaking, the class or interface should depend on a specific model (or model element). Consequently, a change in the model can imply the change of the generated class. Usually, classes that are global for the whole domain are not affected by changes in a model.

We adjusted the category model in Fig. 3a to better fit in with the domain. Fig. 3b shows the modified category model.

The category *A* from Fig. 3a is renamed to *D* (Domain), to emphasize the domain. Consequently, the mixed form *AT* (Application and Technical) becomes *DT* (Domain and Technical). Category *T* remains unchanged. The new category *DG* (domain global) indicates software that is global for the whole domain and helps to differentiate from *D*-classes that are specific to the domain (a particular book, e.g., `CookBook`).

Because of the introduction of *DG*, the characteristic of the *0* category changes somewhat. It contains only global software that is well-tested and *independent of the domain*, e.g., `java.lang` and `java.util` of the JDK. To highlight the difference to the initial *0* definition, *0'* is used.

Figure 4: (a) Addition of software categories (b) Allowed dependencies between categories.

With the above objective in mind and upon searching for generation candidates, in particular, classes of the category *D* are interesting, i.e., *D* itself and *DT*, refining the category of both *D* and *T* (see Fig. 3b).

The matrix in Fig. 4a underscores which software category results if two categories are combined. A usage of *0'* has no effect, e.g., *D + 0' = D*. The same is true for *DG*, as we defined it to be like *0'* (global for the whole domain). Hence, *D + DG = D*, e.g., if the *D*-class CookBook extends the *DG*-class Book it still remains a *D*-class. Only the combination of *D* and *T* leads to an (impure) mixed form, concretely *DT*. Any combination with *DT* results in *DT*, i.e., * + DT = DT.

## 4 DEPENDENCY RULES FOR CATEGORIES

A total of four categories (plus the mixed form *DT*) have been suggested for a general classification of code in generative software (Fig. 3b). Classes of a particular category are only allowed to depend on classes of the same category and classes that are on the same path to *0'*. Consequently, based on these categories, the table in Fig. 4b can be derived automatically.

The table can be read in two ways: line-by-line or column-by-column. The former shows the allowed dependencies *of* a category, whereas the latter shows the categories that may depend *on* a category. The first row in Fig. 4b shows that a *DT*-class may depend on classes of any of the categories. A *D*-class can only depend on *D*-, *DG*- and *0'*-classes[4] (Fig. 4b, second row). A *D*-class must not depend on a *DT*-class. Only the other direction is allowed. Analogous to *D*-classes, a *T*- class may only depend on *T*-, *DG*- and *0'*-classes. A class from category *DG* cannot depend on any of the categories but *DG* and *0'*; otherwise it would contradict the definition of *DG* being global for the whole domain. For example, in the library domain, the (abstract) class Book (*DG*) would not know anything about the single books (such as CookBook,

---

[4]Note that a *D*-class that depends on a *T*-class is rather a *DT*-class.

*D*) or MDDBook (*D*)). Of course, *0'*-classes can only communicate among each other. For instance, classes in the java.lang package (*0'*) do not have any dependencies to a class of any of the other categories.

As mentioned before, the columns in Fig. 4b show those categories that can depend on a specific category. It can be seen that this is somehow antisymmetric to the previously described allowed dependencies of a category.

### Dependencies in Java

Up to now, we included the term dependency, but we did not define it so far. This is mainly because what a dependency ultimately is, depends on the (target) programming language. Java, for example, provides different kinds of dependencies between classes and interfaces. The following shows one possible classification, where the class A depends on the class B and the interface I, respectively:

- Inheritance: class A extends B
- Implementation: A implements I
- Import: import B
- Instantiation: new B()
- ExceptionThrowing: throws B
- Usage: field access (e.g., b.fieldOfB), method call (e.g., b.methodOfB()), declaration (e.g., B b), use as method parameter (e.g., void meth(B b)), etc.

These are dependencies in Java that are mostly manifested in keywords (e.g., extends and throws), and hence, hold for any Java software project. However, not all of these dependencies are always desired. It is important to determine first of all what a dependency ultimately is. For example, an unused import, i.e., a class that imports another class without using it, is not necessarily a dependency.

## 5 CATEGORIZATION APPROACH

The suggested approach for the categorization of the source code is demonstrated in Fig. 5. Three inputs are needed for the categorization: the source code to be categorized (from which a dependency graph is derived), the category graph (such as in Fig. 3b) and an initial categorization of some of the classes and interfaces (usually done by hand). Using these inputs, a categorization tool analyzes the dependencies of the uncategorized classes and interfaces to the already categorized ones. With the information obtained from the category graph some of the uncategorized classes and interfaces can be categorized automatically. For example, if a class C depends on a *D*- and a *T*-class and the category graph in Fig. 3b is given, the category of class C is definitively *DT*, because only this category refines both *D* and *T*.

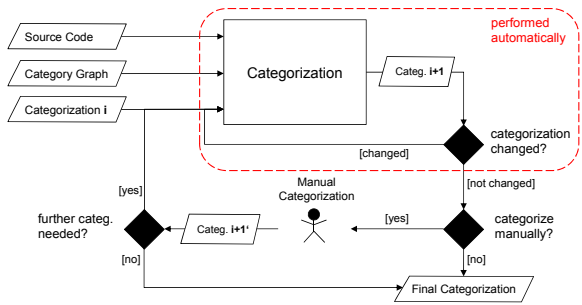

Figure 5: Overview of the categorization approach.

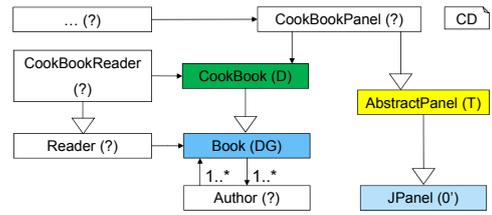

Figure 6: Initially categorized classes.

In some cases the order of the categorization process matters. For example, if a class A only depends on a class B (and no categorized class depends on A), A will not be categorized until B is categorized. To prevent that the order has an effect on the final categorization, the categorization is performed iteratively. The output of iteration *i* serves as input for the next iteration *i+1*. This is repeated until a fixpoint is reached, that means, no further classes and interfaces could be categorized. These iteration steps can be conducted fully automatically. If there are still uncategorized classes left, some of them can be categorized by hand (Sec. 6 illustrates this case by an example). This updated categorization, again can serve as input. The process can be repeated until the whole source code is categorized or no further categorization is needed. Finally, classes and interfaces with a specific categorization serve as candidates for code to be generated. Here, this applies to the categories *D* and *DT*. The user now can decide which of these candidates will become generated code.

## 6 EXAMPLE

Now, with the help of the allowed dependencies defined in Sec. 4, given some classes, the category of each of the classes can be derived semi-automatically, following the approach presented in the previous section.

Consider the case in Fig. 6. The figure depicts overall ten classes, whereby four are pre-categorized (CookBook, AbstractPanel, Book and JPanel) and six are not. The category is in parentheses beside the class name. Uncategorized classes are marked with a question mark (?). Let us assume that the four categorized classes already exist and are categorized (e.g., manually by an expert) and the six other classes are newly created. This situation can arise, for instance, when software evolves. In the following, the categorization process is illustrated.

The class CookBookPanel communicates with both a *D*-class (CookBook) and a *T*-class (AbstractPanel). Following Fig. 4b, only a *DT*-class may communicate with a *D* as well as with a *T* class (marked by a check mark in the *D* and *T* column). Thus, CookBookPanel is definitively a *DT*-class. Moreover, any other class depending on CookBookPanel (represented by the three dots), is also a *DT*-class. In the column *DT* in Fig. 4b there is only a check mark for *DT*.

Next, CookBookReader depends on the *D*-class CookBook and the not yet categorized class Reader. If Reader is a *DT*- or *T*-class, CookBookReader will be definitive a *DT*-class, for it would depend on a *D*-class and either a *DT*- or *T*-class. With regard to Fig. 4b, this only fits for *DT*-classes. If Reader is of any of the other categories, CookBookReader will be a *D*-class. However, when trying to categorize Reader, we encounter a problem. Reader only depends on Book, a *DG*-class. According to Fig. 4b this can apply to any category except *0'*. So, in this iteration, Reader cannot be categorized automatically. Consequently, the exact categorization of CookBookReader cannot be determined.

Analogous to the class Reader, the class Author only depends on the *DG*-class Book. So, except *0'*, it can be of any category. Unlike the previous case, Book also has a dependency to Author, which means that Author is either *DG* or *0'*. We have already excluded *0'*; hence, only *DG* remains as a possible category for Author. Fig. 7 shows the extended categorization after this iteration.

Two classes could not be categorized exactly after the first iteration: CookBookReader and Reader. Recalling that our goal is to find generation candidates, we are above all interested in classes of the category *D*. So, the approximate categorization of CookBookReader (*D* or *DT*) is sufficient, because both *D* and *DT* are of the category *D*. In contrast,

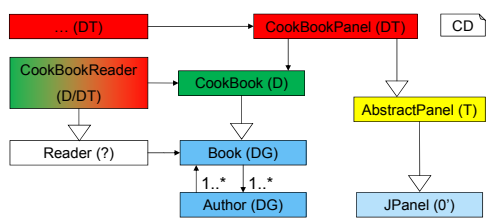

Figure 7: Categorized classes after the first iteration.

`Reader` is still completely uncategorized which hampers the categorization of classes depending on it. There are two options to categorize `Reader` in the next iteration: either manually by the expert or automatically by adding new classes and dependencies limiting the possible categories of `Reader`.

Note that the order of the categorization of `CookBookPanel` and the classes depending on it (marked by "...") is important for the first iteration. The "..." classes could not be categorized if they were considered *before* `CookBookPanel`. However, the order has no impact on the final result, because after the first iteration `CookBookPanel` is surely categorized, and thus, the "..." classes can be categorized in the next iteration.

Finally, three candidates (plus the "..." classes) for generated code are identified: `CookBook` (*D*), `CookBookReader` (*D/DT*) and `CookBookPanel` (*DT*). All of these classes belong to the category *D* directly or indirectly (i.e., *DT*), and hence, are somehow related to the domain. Having these candidates, the generator developer has to decide which of these classes in the end need to be generated and which remain handwritten. Of course, this decision is restricted above all by the information content of the input model. The generator developer must be aware of this restriction.

## 7 FURTHER DEPENDENCIES

Up to now, only the technical dependencies of the code are considered for finding generation candidate classes (see Sec. 4). There can be further dependencies, such as *naming dependencies*. If, for example, the `CookBookPanel` in Fig. 6 had no association to `CookBook`, then, it would only depend on the *T*-class `AbstractPanel` and be a *T*-class.

But, `CookBookPanel` contains the name of `CookBook` as prefix in its class name. Considering this naming dependency, `CookBookPanel` has also a dependency to the *D*-class `CookBook`. Consequently, `CookBookPanel` is a *DT*-class and a generation candidate. Note that from the architecture's point of view a (technical) dependency between `CookBookPanel` and `CookBook` might be forbidden. Hence, deriving the dependency rules from the architecture (and not from software category graph) would limit the kinds of possible dependencies.

In sum, what a dependency finally is, depends on the software system and its conventions. This affects the emerging dependency graph of the source code and can also lead to a different candidate list. However, the procedure as described in Sec. 5 and Sec. 6 remains unchanged.

## 8 CONCLUSION

Code generators are crucial to MDD, transforming abstract models to executable source code. The generated source code often depends on handwritten code, e.g., code from the domain framework. When a code generator is developed or evolved, the generator developer manually decides which classes need to be generated and which remain handwritten. This task can be time-consuming, labor-intensive and may generate more code than is necessary, hampering the maintenance of the software.

This paper has introduced an approach that can aid the generator developer in finding candidates for generated code. First, a software category graph is defined. From this graph the allowed dependencies between the corresponding classes (and interfaces) are derived automatically. After an initial categorization of some classes, further classes can be categorized automatically, by analyzing their dependencies. This procedure is conducted iteratively until all classes are categorized or no more categorization is needed. Finally, generation candidates are all classes belonging to the domain categories.